\documentclass[prl,twocolumn,showpacs,superscriptaddress,preprintnumbers,amsmath,amssymb,groupedaddress]{revtex4}

\usepackage{graphicx}% Include figure files
\usepackage{dcolumn}% Align table columns on decimal point
\usepackage{bm}% bold math
\usepackage{color}% Include figure files
\usepackage{ulem}

\begin{document}

%\preprint{APS/123-QED}
\preprint{}

\title{Hidden quantum 
spin gap state in the static stripe phase of La$_{2-x}$Sr$_{x}$CuO$_{4}$
}% Force line breaks with \\

\author{M. Kofu}
% \altaffiliation[Present address :]{Department of Physics, University of Virginia, Charlottesville, Virginia 22904, USA}%Lines break automatically or can be forced with \\
%\email{mk3ms@virginia.edu}
\author{S.-H. Lee}
\affiliation{
Department of Physics, University of Virginia, Charlottesville, Virginia 22904, USA
}%

\author{M. Fujita}
\affiliation{
Institute for Materials Research, Tohoku University, Sendai 980-8577, Japan
}%

\author{H.-J. Kang}
\affiliation{
NIST Center for Neutron Research, National Institute of Standards and Technology, Gaithersburg, Maryland 20899, USA
}%

\author{H. Eisaki}
\affiliation{
Nanoelectronic Research Institute, National Institute of Advanced Industrial Science and Technology, Tsukuba, Ibaraki 305-8568, Japan
}%

\author{K. Yamada}%
\affiliation{
Institute for Materials Research, Tohoku University, Sendai 980-8577, Japan
}%

\date{\today}% It is always \today, today,

\begin{abstract}
Low energy spin excitations were investigated in the static stripe phase of La$_{2-x}$Sr$_{x}$CuO$_{4}$ using elastic and inelastic neutron scattering on single crystals.  
For $x = 1/8$ in which long-range static stripe order exists, an energy gap of $E_g$ = 4 meV exists in the excitation spectrum in addition to strong quasi-elastic, incommensurate spin fluctuations associated with the static stripes. 
When $x$ increases, the spectral weight of the spin fluctuations shifts from the quasi-elastic continuum to the excitation spectrum above $E_g$. The dynamic correlation length as a function of energy and the temperature evolution of the energy spectrum suggest a phase separation of two distinct magnetic phases in real space. 
\end{abstract}

\pacs{Valid PACS appear here}% PACS, the Physics and Astronomy
\maketitle

%%%%%%%%%%%%%% INTRODUCTION
%\section{Introduction}

After two decades of extensive investigation, some universal characteristics of the magnetic excitations in hole-doped high transition temperature ($T_c$) superconducting cuprates are beginning to emerge. Recent high-energy inelastic neutron scattering studies have shown that, regardless of whether the phase is superconducting or not, YBa$_2$Cu$_3$O$_{6+y}$~(YBCO)~\cite{YBCO_HG}, La$_{2-x}$Ba$_{x}$CuO$_{4}$~(LBCO)~\cite{JM_Tranquada2004} and La$_{2-x}$Sr$_{x}$CuO$_{4}$~(LSCO)~\cite{LSCO_HG, B_Vignolle2007} exhibit strong incommensurate spin fluctuations with an hour-glass type dispersion at high energies: downward and upward dispersing branches of excitations meeting at the $(\pi,\pi)$ point. Several conflicting theories have been proposed to explain the hour-glass excitations that can be grouped into two categories: dynamic stripe models~\cite{CD_Batista2001, GS_Uhrig_etc}  and interacting itinerant fermion liquid models~\cite{DK_Morr1998, MR_Norman_etc}. At the heart of this controversy lies the issue concerning the role of the spin stripes in the mechanism of
 superconductivity. There is now a consensus that the static stripes suppress superconductivity as observed when the hole concentration, $x$, corresponds to the magic number of 1/8 in LBCO and LSCO. When $x$ moves away from 1/8, $T_c$ rapidly increases and the static stripes disappear. On the other hand, the incommensurate spin fluctuations remain strong in the superconducting region of the $x-T$ phase diagram and the incommensurability is proportional to the superconducting phase transition temperature in the underdoped region~\cite{YamadaPRB}. Thus, understanding how the static stripes and incommensurate spin fluctuations are related may provide useful insights towards resolving the controversy. This calls for an investigation of the low energy magnetic excitations in the static stripe phase.

In addition to the formation of static stripes as observed in $x \sim 1/8$~\cite{T_Suzuki1998,H_Kimura1999}, spins respond at low energies by opening a spin gap as observed in $x \sim$ 0.15, the optimal doping concentration~\cite{S_Petit1997, CH_Lee2000}. So far, the two features seem to be exclusive of each other: namely, when static stripes exist, the spin gap does not open and vice versa. Here we show that this is not the case. By performing low-energy inelastic neutron scattering experiments in the static stripe phase of LSCO, we show that even in the presumed static stripe phases ($x=0.125$ and 0.13), a spin gap feature exists in addition to quasi-elastic incommensurate spin fluctuations. 
The spectral weight shift occurs gradually from above $E_g$ to below $E_g$ as $x$ changes from 0.14 to 0.125 while their characteristic wave vectors remain the same. Furthermore, the dynamic correlation length of the spin fluctuations, $\xi$, determined by the inverse of the intrinsic $Q$-linewidth of constant energy scans, is shorter for the excitations above $E_g$, $\xi \sim$ 50~\AA, and is essentially constant for all the concentration considered, than for those below $E_g$ for $x \sim 1/8$, $\xi \sim$ 300~\AA. 
These suggest that there might be a phase separation in the spin stripe phase ($x \sim 1/8$): the superconducting regions with the gapped excitations and the non-superconducting regions of static stripes.

%%%%%%%%%%%%%% EXPERIMENTAL

We have grown several single crystals of LSCO with four different nominal concentrations, $x = 0.125$, 0.13, 0.135, and 0.14, using the traveling solvent floating zone (TSFZ) method~\cite{M_Kofu2005}. Typically, 7 mm diameter and 10 cm long crystals were made for each composition. The crystal was cut into several pieces, two large ones for neutron scattering and small pieces for other characterization measurements. The small pieces from different parts of the original crystal along the rod were used to determine the composition of La, Sr, Cu ions using the inductively coupled plasma (ICP) analysis. The results are listed in Table~\ref{tab:table}. The two large pieces for neutron scattering weighed about 17~g in total for each composition. Bulk susceptibility was measured using a superconducting quantum interference magnetometer~(SQUID). Elastic and inelastic neutron scattering measurements were performed in the ($h$,$k$,0) scattering plane at the cold-neutron triple-axis spectrometer, SPINS, at the NIST Center for Neutron Research and also at the thermal-neutron triple-axis spectrometer, TOPAN, at the Japan Atomic Energy Agency. The experimental configurations for neutron scattering measurements are described in the caption of the figure with the corresponding data. Elastic and inelastic intensities from different compositions were normalized to their sample volumes.

\begin{table}[htb]
\caption{La$_{2-x}$Sr$_{x}$CuO$_4$: Superconducting~(SC) phase transition was observed in the bulk susceptibility measurements. The SC phase transition temperatures $T_{\rm c}$ for four different Sr concentrations, $x$, were  measured either at the mid point or at the onset of the phase transition.}
\vspace{-3mm}
\begin{center}
\begin{tabular}{ccccc}
\hline
nominal $x$ && $x$ & $T_{\rm c}$(midpoint) & $T_{\rm c}$(onset)\\
\hline
$x=0.14$ && 0.130(5) & 36.3~K & 37.1~K\\ 
$x=0.135$ && 0.125(5) & 33.7~K & 35.8~K\\ 
$x=0.13$ && 0.120(6) & 30.9~K & 34.9~K\\ 
$x=0.125$ && 0.115(6) & 27.8~K & 32.1~K\\ 
\hline
\end{tabular}%
\label{tab:table}
\end{center}%
\end{table}%

\begin{figure}[htb]
\begin{center}
\includegraphics[width=\hsize]{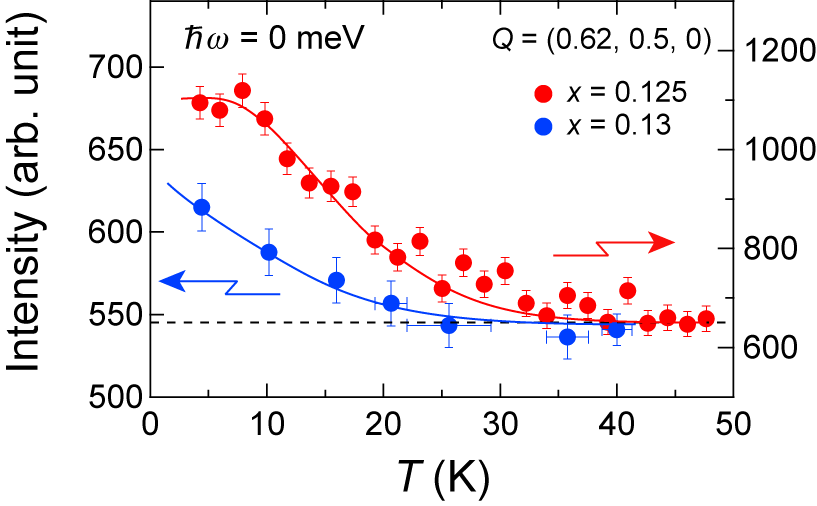}% Here is how to import EPS art
\caption{Temperature dependence of the peak intensity of the elastic incommensurate peak at ${\bf Q} = (0.62,0.5,0)$ for $x = 0.125$ and 0.13. The measurements were performed at SPINS with final neutron energy of $E_f =3.7$~meV and horizontal collimations of Guide-open-80'-open that yielded an energy resolution of 0.18~meV. A liquid nitrogen-cooled BeO filter was placed between the sample and the analyzer in order to eliminate higher order contaminations. The dashed line represents estimated background and the solid lines are guides to the eye.}
\label{fig:elastic}
\end{center}
\end{figure}

\begin{figure}[htb]
\begin{center}
\includegraphics[width=\hsize]{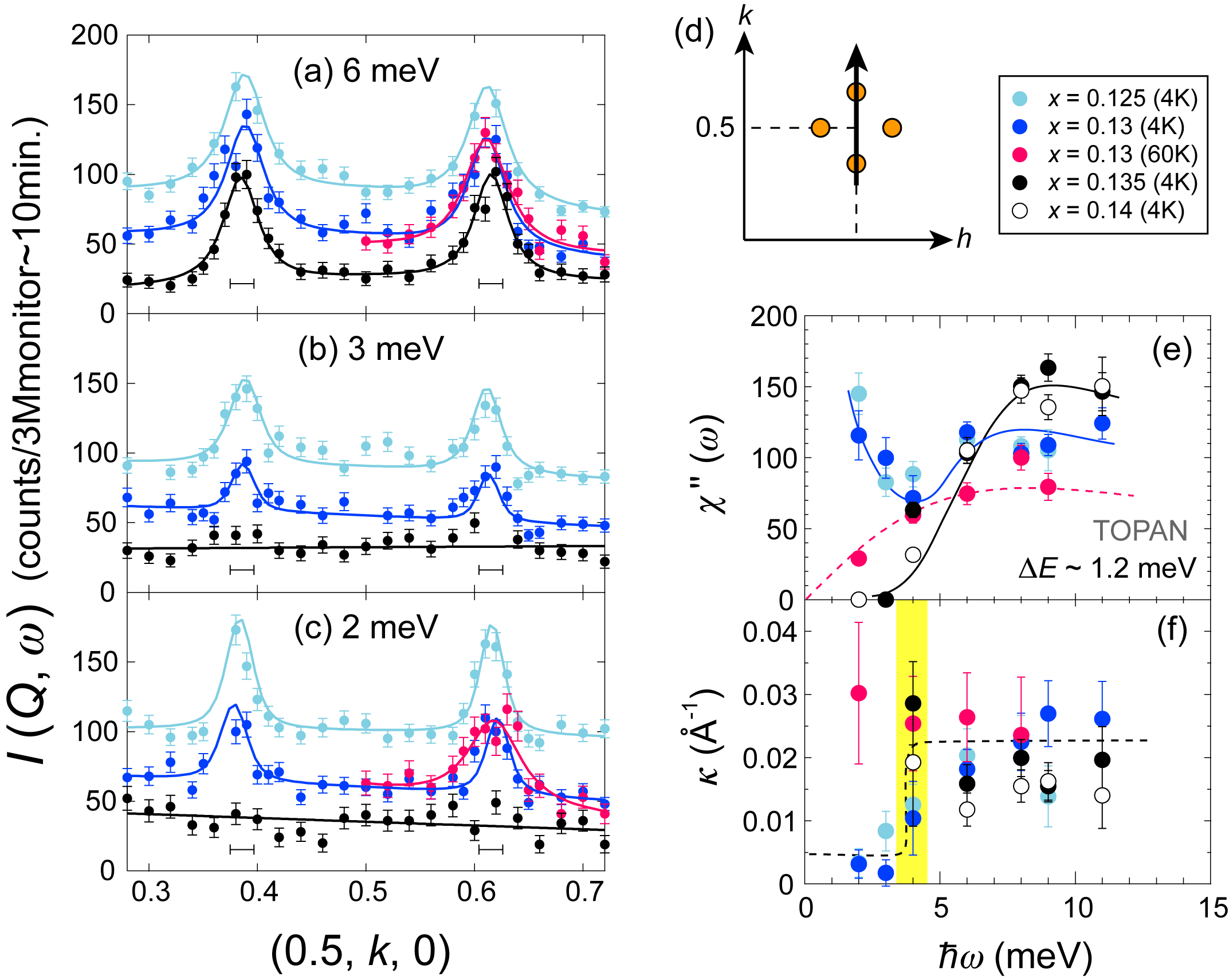}% Here is how to import EPS art
\caption{(a)-(c)~Inelastic constant energy, $\hbar\omega$, scans performed on LSCO ($x =$ 0.125, 0.13, 0.135) with $\hbar\omega =$ 2 meV, 3 meV, and 6 meV, measured at 4 K (black, blue, cyan symbols) and at 60 K (red symbols in (a) and (c)).
Data for $x=0.13$ and 0.125 are shifted for clarity. The solid lines are explained in the text. The measurements were performed at TOPAN with the final neutron energy of $E_f =$ 13.5 meV and horizontal collimations of 40'-30'-30'-180' that yielded an energy resolution of 1.2 meV for $\hbar\omega=3$ meV. A polycrystalline graphite (PG) filter was placed between the sample and the analyzer in order to eliminate higher order contaminations. The horizontal bars represent the resulting intrumental $Q$-resolution. (d)~Momentum space that illustrates the scan direction. (e)~Imaginary part of the spin susceptibility, $\chi ''(\omega )$, and (f)~intrinsic $Q$-linewidth~(half-width-half-maximum), $\kappa$, as a function of the energy transfer, $\hbar\omega$. All data are taken at $T=4$~K except the data for $x=0.13$ taken at $T=60$~K~(red symbols). The solid and dashed lines are guides for eye.}
\label{fig:q-profile}
\end{center}
\end{figure}

\begin{figure}[htb]
\begin{center}
\includegraphics[width=\hsize]{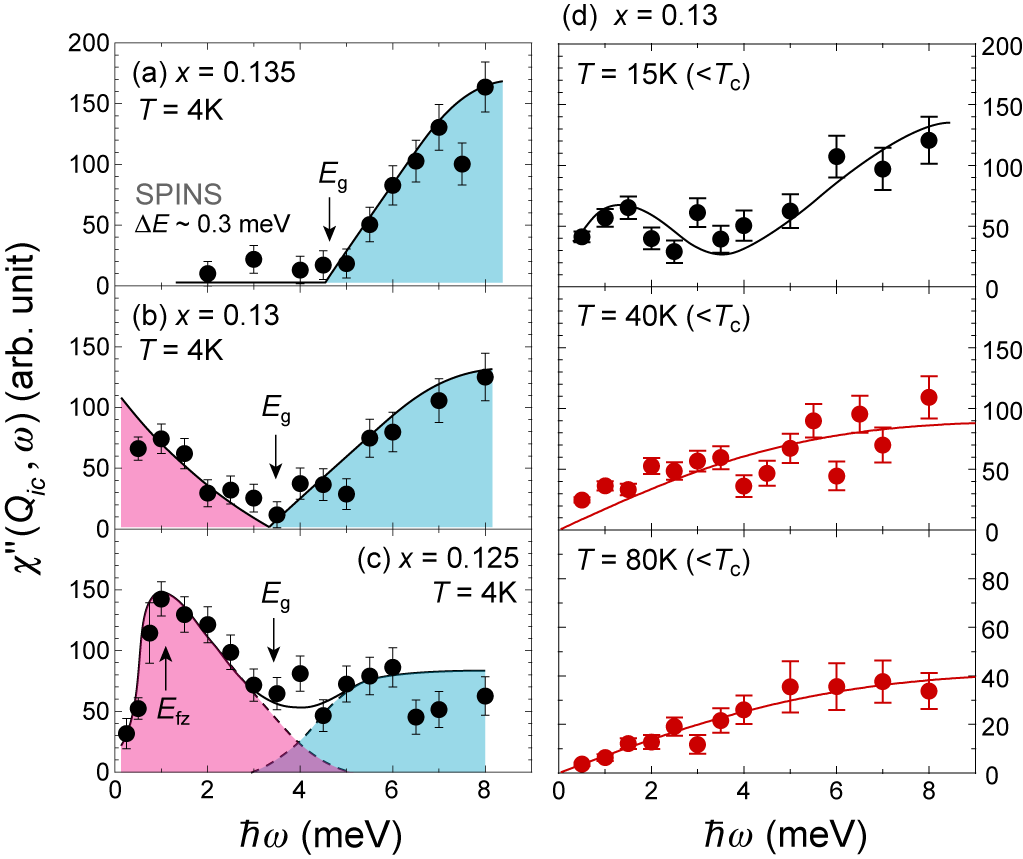}% Here is how to import EPS art
\caption{(a)-(c) 
Energy dependence of $\chi ''(Q_{ic}, \omega )$ for LSCO ($x =$ 0.125, 0.13, 0.135) at 4 K. The measurements were performed using cold-neutrons at SPINS with the same experimental configuration as the one described in the caption of Fig. 1. (d)~Energy spectra for $x=0.13$ at $T=4$, 15, 40, and 80~K.
The solid and dashed lines are guides to the eye.
}
\label{fig:Edep}
\end{center}
\end{figure}

\begin{figure}[htb]
\begin{center}
\includegraphics[width=\hsize]{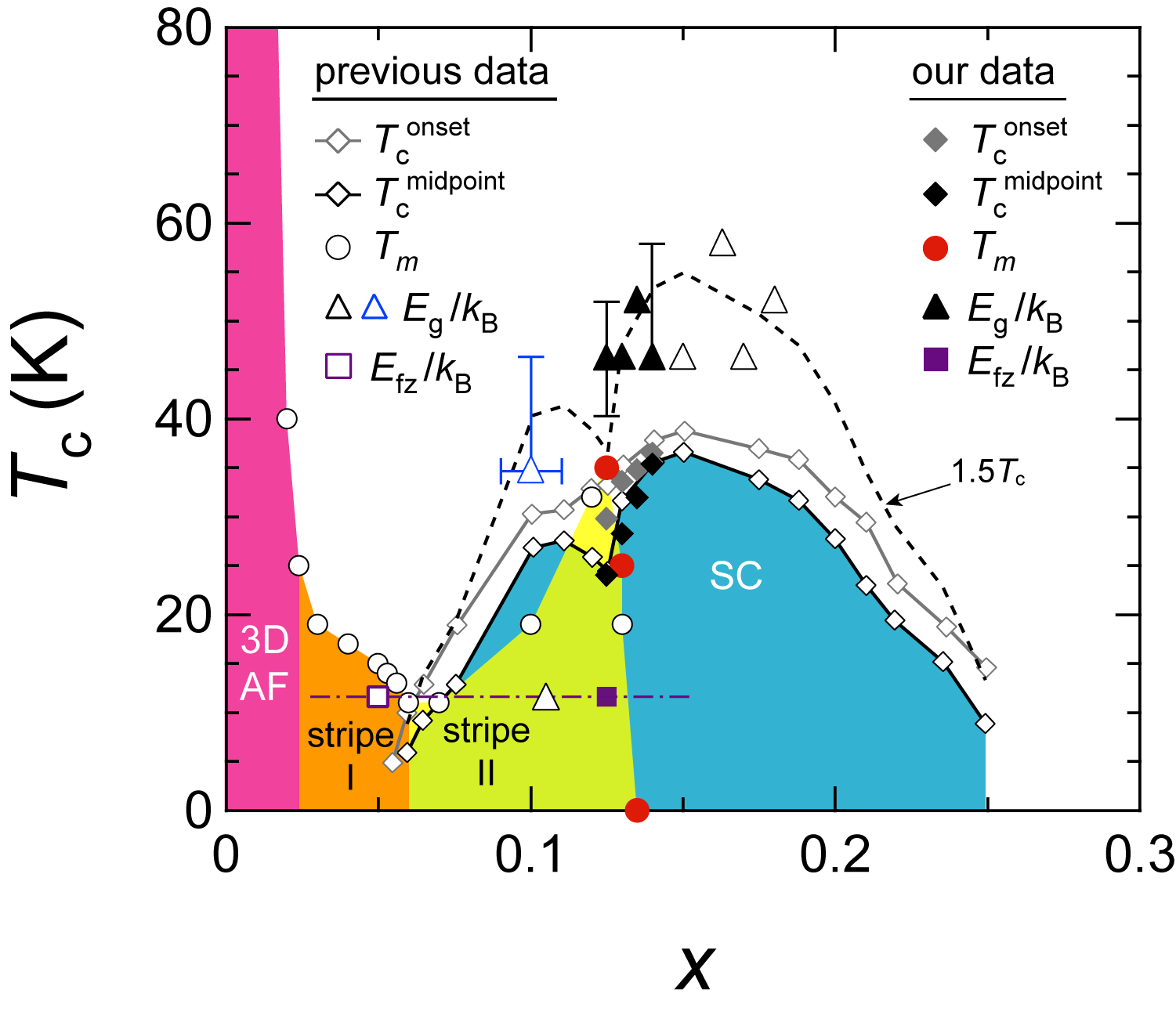}% Here is how to import EPS art
\caption{Phase diagram of La$_{2-x}$Sr$_{x}$CuO$_{4}$ (LSCO) as a function of the nominal hole concentration, $x$, and temperature, $T$. The phase boundaries were obtained by bulk susceptibility ($T_c$) and elastic neutron scattering ($T_m$) measurements on LSCO. Filled symbols are our current data while open symbols are previously reported data for LSCO taken from Ref.~\cite{H_Takagi1989, K_Hirota2001, B_Lake1999, CH_Lee2000, R_Gilardi2004} except the blue open triangle that is for La$_{1.8}$Sr$_{0.14}$Ce$_{0.06}$CuO$_4$ with the hole concentration $x \sim 0.1$\cite{M_Enoki2008}. The stripe I represents the spin-glass phase where static incommensurate peaks appear along the [110] direction that is 45$^\circ$ rotated from the Cu-O bond direction (sometimes called diagonal stripe state). The stripe II represents a stripe phase where static incommensurate peaks appear along the [100] direction as shown in Fig. 2 (d).}

\label{fig:phase}
\end{center}
\end{figure}

%Fig.1 : susceptibility data

As listed in Table~\ref{tab:table}, the superconducting~(SC) transition temperatures
are consistent with previously reported measurements~\cite{H_Takagi1989}. Near $x$ = 1/8, static magnetic incommensurate (IC) peaks appear at low temperatures due to the formation of spin stripes\cite{JMT1995}. Fig. 1 shows that the spin stripe ordering temperature is $T_m =35$~K for LSCO ($x = 0.125$). When $x$ moves away from 1/8, the spin order appears at a lower temperature, $T_m =25$~K for $x = 0.13$, and the peak intensity weakens (see blue circles in Fig. 1).  By $x = 0.135$, the peak becomes undetectable. Previous neutron scattering studies have reported that when the hole concentration is close to the optimal doping $x \sim 0.15$, the static order disappears and a spin gap opens up at $\hbar\omega \sim 4$~meV~\cite{S_Petit1997, CH_Lee2000}.  

In order to investigate how the two features, the static stripe and the spin gap state, evolve with the hole concentration, $x$, we have systematically performed low energy inelastic neutron scattering measurements on La$_{2-x}$Sr$_{x}$CuO$_{4}$ with the four different Sr concentrations from $x$ = 0.125 to $x$ = 0.14. Fig.~2 shows constant energy ($\hbar\omega$) scans performed as a function of momentum transfer along the (0.5,$k$,0) direction as shown in Fig. 2 (d). For $\hbar\omega =$ 6 meV (Fig.~2~(a)), strong inelastic IC spin fluctuations are present in all concentrations. 
When the energy transfer is lowered to $\hbar\omega =3$~meV and 2~meV (Fig.~2~(b) and (c)), the spin fluctuations are still strong for $x = 0.125$ and 0.13 that have the static spin stripe order. For $x = 0.135$ when the static order is absent, however, the low energy spin fluctuations become very weak. This suggests an opening of a spin gap for $x = 0.135$ at $E_g \sim$ 3~meV, as expected since $x$ is approaching the optimal $x = 0.15$~\cite{S_Petit1997, K_Yamada1995, CH_Lee2000}. More interesting in Fig.~2 is that even for $x = 0.125$ and 0.13 the spin fluctuations are weaker for $\hbar\omega =3$~meV than for $\hbar\omega=2$~meV and 6~meV, that will be discussed below in more detail.

We have performed similar constant-$\hbar\omega$ scans at 4 K with several different $\hbar\omega$ values up to 11 meV in order to study the low energy dependence of the IC spin fluctuations.
The data were fit to the four scattering rods stemming from the four IC reflections convoluted with the instrumental resolution function to extract the imaginary part of the $Q$-integrated spin susceptibility, $\chi ''(\omega )$ and the intrinsic $Q-$linewidth, $\kappa$. Fig.~2~(e) and (f) show the results. Fig.~2~(e) shows that for $x = 0.135$ and 0.14, there is a spin gap of $\sim 3$ meV below which no magnetic scattering was observed. For $x$ = 0.13 and 0.125, the spin gap is filled but there seems to be a dip at $\hbar\omega \sim 4$~meV. In order to investigate more carefully the dip feature in the energy spectrum, we have performed inelastic neutron scattering measurements with a sub-meV energy resolution available at the cold-neutron triple-axis spectrometer, SPINS. Fig. 3 (a)-(c) shows constant-$\bf Q$ scans measured at ${\bf Q} = {\bf Q}_{ic}$ and at 4 K as a function of $\hbar\omega$ for $x = 0.125$, 0.13 and 0.135. It is clear that when there is no static IC order for $x = 0.135$ the spectrum has a spin gap of $E_{g} \sim 4.5$ meV above which $\chi''({\bf Q}_{ic},\hbar\omega)$ increases with increasing $\hbar\omega$ (see Fig.~3~(a)). When there is static IC order in $x = 0.125$ and 0.13, on the other hand, additional IC spin excitations appear below the gap. A similar low energy spin gap feature has also recently been observed in La$_{1.8}$Sr$_{0.14}$Ce$_{0.06}$CuO$_4$~\cite{M_Enoki2008}. 
Two features should be noted here: (1)~Even in the static spin stripe phase ($x = 0.13$ and 0.125) the energy spectrum has a dip at $E_{g} \sim 4$~meV at which $\chi''({\bf Q}_{ic},\hbar\omega)$ is weak (see Fig.~3~(b) and (c))~\cite{footnote}. (2)~The low energy IC spin excitations below $E_{g}$ appear at the expense of those above $E_{g}$. 
As shown by filled triangles in Fig. 4, our $E_{g}$ values divided by the Boltzmann constant, $k_{\rm B}$, are close to the previously determined spin gaps for higher doping concentrations (open triangles), all of which seem to be close to $E_{g}= 1.5 k_{\rm B} T_c$ (dashed line). For comparison, $E_{g} \sim 3.5 k_B T_c$ for YBCO~\cite{YBCO,note}. Further studies near the critical concentration ($x_c = 0.06$) are necessary in order to clarify the relationship between $E_g$ and $T_c$ in LSCO.

What is the origin of the dip in the energy spectrum of $x\sim 1/8$? The intrinsic $Q$-linewidth, $\kappa$, as a function of $\hbar\omega$, that is inversely proportional to the dynamic spin correlation length, $\xi$ can provide important information on this issue. As shown in Fig. 2 (f), at 4 K $< T_c$, $\kappa \sim$ 0.02 \AA$^{-1}$ (or $\xi \sim$ 50 \AA) for $\hbar\omega \geq 4$ meV for all concentrations considered, while it is much shorter, $\kappa \sim 0.003$ \AA$^{-1}$ (or $\xi \sim$ 300  \AA) for $\hbar\omega \leq 3$~meV for the static stripe phase ($x = 0.13$ and 0.125). 
This suggests that the low energy spin fluctuations that exist below the energy dip ($E_g$) for $x \sim 1/8$ have different characteristics than the fluctuation above the dip. We also studied how the low and high energy excitations of the $x\sim 1/8$ samples evolve upon warming by performing the energy scan on the $x$ = 0.13 sample at various temperatures. As shown in Fig. 3 (d), the dip feature in the energy spectrum seems to survive at 15 K $< T_m \sim T_c$ at which the static spin stripe exists, while for $T > T_c$ at which the static spin stripe is melted the dip is now replaced by a single lorentzian behavior as expected for a paramagnetic phase. Furthermore, $\kappa$ measured at 60 K $> T_m \sim T_c$ becomes nearly constant for all $\hbar\omega$ up to 8 ~meV (red symbols in Fig. 2 (f)). These suggest that the dip feature may be an indication of the existence of two magnetic phases in LSCO ($x\sim 1/8$) that are distinctly separated in real space.

There is an additional feature in the energy spectrum of x = 0.125 (Fig. 3 (c)): a dip below $E_{fz} \sim 1$ meV. This is not uncommon in a system that undergoes a spin freezing upon cooling that a static central mode appears  and very low energy spin excitations become damped. Such behaviors have been observed in cuprates such as $E_{fz} \sim 1$~meV in the spin glass phase of LSCO ($x=0.05$)~\cite{W_Bao2007} and $E_{fz} \simeq 2$~meV in the SC phase of YaBa$_2$Cu$_3$O$_{6.353}$~\cite{C_Stock2006}. Recently, similar behavior was also observed in LSCO ($x=0.105$) and was interpreted as $E_{g}$ related to the spin gap of the optimal concentration.~\cite{J_Chang2007}
As shown in Fig.~4, however, the gap value of 1~meV (the open triangle at $x = 0.105$) lies on the dotted dashed line connecting the two $E_{fz}$ values for $x = 0.05$ and 0.125, which suggests that the gap is associated not with superconductivity but with $E_{fz}$ observed for $x \sim 1/8$ and in the spin glass phase. This interpretation is supported by a very recent work on LBCO ($x=1/8$)~\cite{JS_Wen2008} that showed the $\sim$1 meV gap is insensitive to the external magnetic field.

Our neutron scattering data showed that for LSCO ($x\sim 1/8$) a static stripe order appears along with low energy spin excitations below the spin gap at the expense of spin fluctuations above the gap, leaving the spin gap intact. The features observed in the spin excitation spectrum above and below $E_{g}$ have the same characteristic incommensurate wave vector, and the gradual shift in their spectral weight with changing $x$ indicates that the two features have the same origin.
% In other words, the high energy hour-glass excitations above $E_g$ are most likely due to dynamical stripes~\cite{GS_Uhrig_etc} that can partially freeze into the static stripes when $x \sim 1/8$. Some questions arise from this scenario: (1)~how do the static spin stripes appear in real space? and (2)~why do the hour-glass excitations have a spin gap when the system is superconducting?
On the other hand, our finding that the intrinsic $Q$-linewidth, $\kappa$ dramatically changes at $E_g$ in the static stripe phase
suggests a real space phase separation of two distinct magnetic phases: the superconducting regions with gapped spin fluctuations and nonsuperconducting regions with static spin stripes. Another scenario can be that low frequency motions of the domain walls between the spin stripes may be responsible for the observed exotic low energy spin fluctuations.\cite{zaanen96} A recent $\mu$SR study suggested that in cuprates there are two distinct phases separated in real space: regions with fluctuating moments and other regions with frozen moments, and their volume fractions change with $T_c$~\cite{AT_Savici2005}, supporting the first scenario of phase separation. However, the exact origin of the spin gap feature in LSCO ($x\sim 1/8$) could be fully understood by further studies, such as magnetic field effects on the spin fluctuations below and above $E_g$.

%\rev{In addition, neutron scattering studies show that an additional low-energy excitation is induced by applied magnetic field or Zn substitution~\rev{\cite{B_Lake2001,  M_Kofu2005, H_Kimura2003}} that make non-superconducting~(non-SC) islands in the superconducting sea.
%The additional slow fluctuation is induced in the non-SC region.}
%As for the spin gap in the superconducting cuprates, several theories have shown that $T_c$ increases when the spectral weight of spin fluctuations shifts from low to higher energies~\cite{P_Monthoux_etc}, that means that superconductivity might push the low energy spin excitations up in energy to form a spin gap in the hour-glass spectrum. When SC disappears such as in LBCO ($x=1/8$), the gapped hour-glass excitations are replaced by gapless hour-glass excitations.

%%%%%%%%%%%%%% DISCUSSIONS

%%%%%%%%%%%%%% Conclusion

%%%%%%%%%%%%%% ACKNOWLEDGMENTS
%\begin{acknowledgments}
We thank J. M. Tranquada, Y. B. Kim,  M. Matsuura, T. Matsumura for helpful discussions, C. H. Lee, K. Hirota and T. J. Sato for their support during sample growth and bulk property characterization. Work at University of Virginia was supported by the US DOE through DE-FG02-07ER46384 and the US DoC-NIST through 70NANB7H6035.

%Activities at UVA is supported by ...
%\end{acknowledgments}

%The neutron scattering experiment was performed under the collaboration in neutron science between High Energy Accelerator Research Organization 
%and Los Alamos National Laboratory. 
%This work was supported by a Grant-in-Aid for Creative Scientific Research (No. 16GS0417) from the Ministry of Education, Culture, Sports, Science and Technology of Japan.

%\newpage %Just because of unusual number of tables stacked at end
%\bibliography{ref}% Produces the bibliography via BibTeX.

\begin{thebibliography}{50}
\bibitem{footnote} The dip in the energy spectrum is very clear for $x=0.13$ (Fig. 3 (b)) while it is less obvious for $x=0.125$ (Fig. 3 (c)). However, as shown in Fig. 2 (f) $\kappa$ shows the same sharp change in energy at $\sim 4$ meV for both concentrations, which we believe manifests the existence of the dip even in the $x=0.125$ compound.
\bibitem{note} Note that for YBCO $E_{g}$ was determined as the mid-point of the upturn of the intensity~\cite{YBCO} while for LSCO $E_g$ is the energy above which the excitations appear.
\end{thebibliography}

\end{document}